\documentstyle[12pt,multibox]{article}
\newtheorem{lemma}{Lemma}
\begin{document}
\begin{titlepage}

\begin{flushright}

ULB--TH--98/13\\
hep-th/9807016\\
July 1998\\

\end{flushright}
\vfill

\begin{center}
{\Large{\bf  Couplings between generalized gauge fields}}
\end{center}
\vfill

\begin{center}
{\large
J. Antonio Garcia\,$^{a}$\ and
Bernard Knaepen\,$^{b}$}
\end{center}

\begin{center}{\sl
$^a$ Instituto de Ciencias Nucleares
Universidad Nacional Aut\'onoma de M\'exico
Apartado Postal 70-543, 04510 M\'exico, D.F.
\\
$^b$ Facult\'e des Sciences, Universit\'e Libre de Bruxelles,
Campus Plaine C.P. 231, B--1050 Bruxelles, Belgium.

}\end{center}
\vfill

\begin{abstract}
We analyze the BRST field-antifield construction 
for generalized gauge fields consisting of massless mixed
representations of the Lorentz Group and we calculate
all the strictly gauge invariant interactions between them. All
these interactions are higher derivative terms constructed out
from the derivatives of the curl of field strength.
\end{abstract}
\vfill
\vspace{1cm}
\hrule width 5.cm
\vspace{.2cm}

\footnotesize \noindent PACS: 2.20.-a, 11.15.-q, 11.10.Kk,
11.25.-w
\end{titlepage}
\normalsize
\section{Introduction}
Generalized gauge fields consisting of 
free massless integer spin mixed representations of the 
Lorentz group (mixed tensors) can be used as models for higher
spin covariant bosonic theories and to
study the long standing problem of interactions for massless higher
spin particles. A geometric formulation for this type of
theories is still an open question.

Another source of interest for this type of theories  comes
from the formulation of D=11 dimensional supergravity as a gauge theory
for the $osp(32|1)$ superalgebra. 
In addition to the vielbein $e^a_\mu$, the spin
connection $\omega^{ab}_\mu$ and the gravitino $\psi_\mu$ the
connection for $osp(32|1)$ contains a totally antisymmetric
fifth rank Lorentz tensor one form $b_\mu^{abcde}$
\cite{Holten,Towsend,Zanelli}.  The antisymmetric part of
this tensor could be identified to an abelian six form
$A_{[6]}$
\cite{Nicolai,Zanelli} 
but the mixed tensor
part does not have any related counterpart
in the standard D=11 supergravity theory. 

Free consistent massless theories
whose field content is this class of mixed tensors have
attracted some attention in the past. In particular,
possible lagrangians, compatible with gauge invariance, have
been proposed in \cite{Aulakh1,Chung1,Curtright1} while the
complete ghost spectrum and the BRS operator are described in 
\cite{Aulakh1,Chung1,Labastida1}.

In this article we extend the BRS analysis to include {\em
antifields} and calculate the cohomology of the
``longitudinal exterior derivative" in order to obtain all
the consistent, strictly gauge invariant interactions that can
be added to the free theory.

{}For simplicity, we will concentrate on tensors
with three  indices which satisfy the following identities:
\begin{equation}
T_{[ab]c}=-T_{[ba]c},\quad T_{[ab]c}+T_{[ca]b}+T_{[bc]a}=0;
\label{defT}
\end{equation}
we comment at the end of the paper on how to generalize our
results to other mixed tensors. In terms of Young diagrams
the fields (\ref{defT}) are represented by
\
\begin{picture}(20,22)(0,0)
\multiframe(0,10)(10.5,0){2}(10,10){a}{c}
\multiframe(0,-0.5)(10.5,0){1}(10,10){b}
\end{picture}.

The paper is organized as follows.  First we present the
analysis of the model within the Hamiltonian formalism
and we rederive some of the results obtained in \cite{Curtright1}.
Then we present the BRST
field-antifield formalism for the theory.  After that we obtain
the consistent vertices of the theory by calculating the
cohomology of the exterior longitudinal derivative.  Finally, we
discuss how the method can be applied to other higher rank
Lorentz tensors corresponding to more complicated Young diagrams.

\section{Hamiltonian formalism}
Using the `Hook' formula \cite{Dragon1,hamermesh}, one
easily calculates that the tensors (\ref{defT}) have
$\frac{1}{3}D(D-1)(D+1)$ components in $D$ dimensions.

The lagrangian of the theory is,
\begin{equation}
{\cal L}=-\frac{1}{12}\left(F_{[abc]d}F^{[abc]d}-3F_{[abx]}^{\ \ \ \ x}
F_{\ \ \ \ \ y}^{[aby]}\right),
\label{Lagrangian}
\end{equation}
where $F_{[abc]d}=\partial_{a}T_{[bc]d}+\partial_{b}T_{[ca]d}
+\partial_{c}T_{[ab]d}$
and the corresponding action is invariant under the gauge
transformations,
\begin{equation}
\delta_{\epsilon,\eta}T_{[ab]c}=\partial_a \epsilon_{bc}-\partial_b
\epsilon_{ac}+\partial_a \eta_{bc}-\partial_b \eta_{ac}-2\partial_c
\eta_{ab},
\label{invdejauge}
\end{equation}
where $\epsilon_{ab}$ are symmetric and $\eta_{ab}$ are antisymmetric
gauge parameters.
The Euler-Lagrange equations are,
\begin{equation}
	\frac{\delta {\cal L}}{\delta T_{[ab]c}}=\frac{1}{2}E^{[ab]c}-
\frac{1}{4}(g^{bc}E^a-g^{ac}E^b)=0,
\label{ELeq}
\end{equation}
where $E_{[ab]c}=\partial^dF_{[abd]c}-\partial_c
F_{[abx]}^{\ \ \ \ x}$ and $E_a=E_{[ax]}^{\ \ \ x}=2\partial^y
F_{[yax]}^{\ \ \ \ x}$.
By taking the trace of (\ref{ELeq}) we see that the
equations of motion can be equivalently written,
\begin{equation}
	E_{[ab]c}=0,\quad E_a=0.
\end{equation}
These equations of motion satisfy the Noether identities,
\begin{equation}
\partial_a E^{[ab]c}
+\frac12\partial^c E^b \equiv 0, \quad \partial_a E^a \equiv 0.
\end{equation}

To proceed with the Hamiltonian analysis we write
(\ref{Lagrangian}) as,
\begin{eqnarray}
{\cal L}&=&-\frac{1}{12}(3 F_{[0ij]k}F^{[0ij]k}+ F_{[ijk]d}F^{[ijk]d}
\nonumber\\
&&\hspace{0cm}- 6 F_{[ij0]}^{\ \ \ \ 0} F^{[ijk]}_{\ \ \ \ k}
-6 F_{[0ik]}^{\ \ \ \ k} F^{[0il]}_{\ \ \ \ l}
-3 F_{[ijk]}^{\ \ \ \ k} F^{[ijl]}_{\ \ \ \ l}).
\end{eqnarray}
Therefore the momenta are given by,
\begin{equation}
\pi^{[ij]k}=\frac{\partial{\cal L}}{\partial\partial_0 T_{[ij]k}}=
-F^{[0ij]k}+g^{jk}F^{[0il]}_{\ \ \ \
l}-g^{ik}F^{[0jl]}_{\ \ \ \ \
l},
\end{equation}
\begin{equation}
\pi^{[ij]}_{\ \ \ 0}=\frac{\partial{\cal L}}{\partial\partial_0
T_{[ij]}^{\ \ 0}}= F^{[ijk]}_{\ \ \ \ \ k},
\end{equation}
where we have used the convention, $\frac{\partial T^{[ab]c}}
{\partial T^{[de]f}}=
(\delta^a_d\delta^b_e-\delta^a_e\delta^b_d)\delta^c_f $.
The theory has $(D-1)^2+(D-1)$ primary constraints,
namely: $\pi^{[0i]0}=0$ and
$\pi^{[0i]k}=0$.

The canonical Hamiltonian is then given by,
\begin{eqnarray}
H_c&=&\frac12 \pi^{[ab]c}\partial_0 T_{[ab]c}-{\cal L}
\\ &=& \frac14 \pi^{[ij]k}\pi_{[ij]k}-\frac{1}{2(D-3)}
\pi^{[il]}_{\ \ \ l}
\pi_{[ik]}^{\ \ \ k}\nonumber+
\frac{1}{12} F_{[ijk]d} F^{[ijk]d}\\&&-
\frac14 F_{[ijk]}^{\ \ \ \ k} F^{[ijl]}_{\ \ \ \ l}
\nonumber + T_{[j0]}^{\ \ \ 0}\partial_i \pi^{[ij]}_{\ \ \ 0}+
 T_{[j0]k}\partial_i \pi^{[ij]k}.
\label{Hcanon}
\end{eqnarray}
Using the Poisson-Bracket,
\begin{equation}
\{T_{[lm]n}(x), \pi^{[ij]c}(y)\}=
\delta_n^c(\delta^i_l\delta^j_m-\delta_m^i\delta^j_l)\delta(x-y),
\end{equation}
one easily obtains the secondary constraints:
\begin{equation}
\partial_i \pi^{[ij]k}=0, \quad
\partial_i \pi^{[ij]}_{\ \ \ 0}=0.
\end{equation}
These can also be read out from (\ref{Hcanon}) since
$T_{[j0]}^{\ \ \ 0}$ and $T_{[j0]k}$ play the role of Lagrange
multipliers. The number of secondary class constraints is
the same as the primary constraints. However the former
provide a reducible set because we have identically,
\begin{equation}
\partial_j\partial_i \pi^{[ij]k}\equiv0, \quad
\partial_j\partial_i \pi^{[ij]}_{\ \ \ 0}\equiv 0.
\label{reducrel}
\end{equation}
The total number of effective constraints is thus
$2((D-1)^2+(D-1))-D=D(2D-3)$ and therefore the theory
possesses $2\frac13 D(D-1)(D+1)-2D(2D-3)=\frac13
D(D-2)(D-4)$ degrees of freedom (all the constraints are first
class). Note that in 4 dimensions the theory is therefore
trivial.

The identities (\ref{reducrel}) induce reducibility
identities among the gauge transformations and are
responsible for the presence of ghosts of ghosts in the BRS
ghost spectrum. Indeed, the gauge variations vanish
for the choices,
\begin{equation}
\epsilon_{ab}= 3(\partial_a C_b+\partial_b C_a),
\end{equation}
\begin{equation}
\eta_{ab}= \partial_a C_b-\partial_b C_a,
\end{equation}
where $C_a$ are $D$ arbitrary functions. 
\section{BRST field-antifield formalism}
According to the general rules of the BRST field-antifield
formalism \cite{HenneauxTeitelboim}, the BRST differential
$s$ is constructed as follows.

First, one defines a differential $\delta$ called the
Koszul-Tate differential whose role is to implement the
equations of motion in cohomology. Therefore, we first
introduce the antifields $T^{*[ab]c}$ which satisfy,
\begin{equation}
\delta T^{*[ab]c}=-\frac{\delta {\cal L}}{\delta T_{[ab]c}}=
-\frac{1}{2}E^{[ab]c}+
\frac{1}{4}(g^{bc}E^a-g^{ac}E^b).
\label{anti1}
\end{equation}
Because of the presence of Noether identities among the
Euler-Lagrange equations, we also need the
antifields $T^{*bc}$ which satisfy,
\begin{equation}
\delta T^{*bc}=\partial_a T^{*[ab]c}.
\label{anti2}
\end{equation}
Finally, we introduce the antifields $T^{*c}$ which satisfy,
\begin{equation}
\delta T^{*c}=\partial_b T^{*bc}.
\label{anti3}
\end{equation}
These are present because the theory is reducible.
The roles of $T^{*bc}$ and $T^{*c}$ are respectively to eliminate
from the cohomology of $\delta$ the terms
$\partial_a T^{*[ab]c}$ and $\partial_b T^{*bc}$
which would otherwise be present.
Each antifield carries a degree called the `antighost'
number which is given
by, $antighost(T^{*[ab]c})=1$, $
antighost(T^{*bc})=2$, $antighost(T^{*c})=3$ and
$antighost(\delta)=-1$.

Because of the definitions
(\ref{anti1}), (\ref{anti2}) and (\ref{anti3}),
the cohomology of $\delta$ in the algebra generated by the
fields, the antifields and their derivatives is given by the
on-shell functions; in other words, $\delta$ provides a
resolution of the stationary surface.

More important to us in the BRST construction is the
longitudinal exterior derivative $\gamma$ which takes into
account the gauge invariance of the model. In our case, we
first need to introduce the ghosts $S_{ab}$ and $A_{ab}$
in place of each gauge
parameter according to the definition,
\begin{equation}
\gamma T_{[ab]c}=\partial_a S_{bc}-\partial_b
S_{ac}+\partial_a A_{bc}-\partial_b A_{ac}-2\partial_c
A_{ab},
\end{equation}
where $S_{ab}$ and $A_{ab}$ are respectively symmetric and
antisymmetric in $ab$.

Because the gauge transformations are reducible we also need
the ghosts of ghosts $C_a$ which satisfy,
\begin{eqnarray}
\gamma S_{ab}=3(\partial_a C_b + \partial _b C_a), \label{varG1}\\
\gamma A_{ab}=\partial_a C_b - \partial _b C_a.
\label{varG2}
\end{eqnarray}
With these definitions, we have $\gamma^2=0$.
A grading called the `pureghost' number  is associated to the ghost
fields and we
have: $pureghost(S_{ab})=pureghost(A_{ab})=1$,
$pureghost(C_a)=2$ and also $pureghost(\gamma)=1$. Note that the
fields and their derivatives are of antighost and pureghost
number 0 and that $\gamma(antifields)=\delta(ghosts)=0$.

{}For the model we consider, the full BRST differential is
simply given by the sum of the Koszul-Tate differential and
the longitudinal exterior derivative: $s=\delta +\gamma$.
The grading of $s$ is called the `ghost' number and is
given by $ghost=pureghost-antighost$.

It is obvious from the above definitions that a combination
of the fields and their derivative will be strictly gauge invariant
if and only if it defines an element of the cohomology
$H(\gamma)$ of the differential $\gamma$.

It was shown in \cite{BH1} that the classification of all
the consistent interactions that can added to a free
action can be obtained by calculating in ghost number zero
the cohomology $H(s\vert d)$ of the BRST differential $s$. These
consistent interactions can be grouped in three categories.
In the first one, we have the vertices which are strictly
gauge invariant; their study only requires the calculation
of the $\gamma$-cohomology. The second category consists of
interactions
which are gauge invariant up to a boundary term. Finally the
last category contains the vertices which are gauge
invariant up to a boundary term on-shell. These last
interactions therefore require a modification of the gauge
transformations and correspond to antifield dependent
elements of the BRST cohomology whereas the first two
categories correspond to antifield independent solutions
of $H(s \vert d)$.

In this article we calculate all the consistent interactions
of the first category, i.e, the gauge invariant functions.

\section{Gauge invariant functions}
In this section we obtain our main result, namely, we
calculate all the gauge invariant terms that can be added
to the free lagrangian.
As we recalled, these are polynomials in the fields and
their derivatives which belong
to the cohomology of the differential $\gamma$. However,
because its study is important in order to obtain the other
consistent interactions \cite{henneauxSC}, e.g. to solve the
`descent equations', we will calculate $H(\gamma)$
in  the full algebra generated by the fields, the ghosts, the
antifields and their derivatives.

The procedure we use is based on the following result,
sometimes referred to as the ``Basic Lemma"
\cite{HenneauxTeitelboim}:
\begin{lemma}
Let ${\cal A}$ be the polynomial algebra generated by the
algebraically independent variables
$x^i$, $y^{\alpha}$, $z^{\alpha}$ and let $D$ be a differential
whose action on the variables is:
\begin{equation}
D x^i=0, \quad D y^{\alpha}= z^{\alpha}.
\label{regroup}
\end{equation}
The cohomology of $D$ in ${\cal A}$,
 $H(D)\equiv \frac{Ker \ D}{Im \ D}$, is
then given by the polynomials in the $x^i$.
\end{lemma}
In our case the algebra ${\cal A}$ is the algebra generated
by the fields, the antifields, the ghosts, the ghosts of ghosts and all
their derivatives. Our task is thus to redefine all our
generators in such a way that they obey (\ref{regroup}).

First of all, let us note that the antifields and their
derivatives are all
$\gamma$-closed and do not appear in the $\gamma$
variations of the other fields. This implies that they are
automatically part of the $x^i$ variables.

For the other variables, we denote by $V^k$ the vector space
spanned by $\partial_{s_1 \ldots s_k}C_a$, $\partial_{s_1 \ldots
s_{k-1}}A_{ab}$, $\partial_{s_1 \ldots s_{k-1}}S_{ab}$,
$\partial_{s_1 \ldots s_{k-2}}T_{[ab]c}$. Our hole
algebra is ${\cal A}=\oplus_k V^k$ and $\gamma$ has a well
defined action in each $V^k$. We will therefore look for new
coordinates in each
$V^k$.

In $V^0$ we have,
\begin{equation}
\gamma C_a =0.
\label{gammaCa}
\end{equation}
$C_a$ is therefore a variable of type $x^i$.

In $V^1$ the variables split as in (\ref{varG1}),
(\ref{varG2}). $S_{ab}$ and $A_{ab}$ are therefore of type
$y^\alpha$ while the symmetrized and antisymmetrized first
order derivatives of the $C_a$ are of type $z^\alpha$.

For the higher order $V^k$, the analysis proceed in the
same fashion but the algebra becomes more involved. The
calculus is simplified by the systematic use of Young
diagrams in order to decompose the variables into
irreducible parts.

Let us first examine the fields and their derivatives. Because
partial derivatives commute, the $k$-th
order derivatives of the fields can be symbolically
represented by the following tensor product of Young
diagrams: $\partial_{d_1 \ldots d_k}T_{[ab]c} \equiv$
\footnotesize \begin{picture}(106,25)(0,0)
\multiframe(0,0)(10.5,0){6}(10,10){$d_1$}{.}{.}{.}{.}{$d_k$}
\put(67.5,3){$\bigotimes$}
\multiframe(82,10)(10.5,0){2}(10,10){$a$}{$c$}
\multiframe(82,-0.5)(10.5,0){1}(10,10){$b$}
\end{picture}.\normalsize

According to the general theory of the representations of
the symmetric group, this tensor product decomposes into the
following irreducible components under a general invertible
transformation,

\footnotesize \begin{picture}(210,100)(0,-60)
\multiframe(0,0)(10.5,0){6}(10,10){$d_1$}{.}{.}{.}{.}{$d_k$}
\put(67.5,2){$\bigotimes$}
\multiframe(82,10)(10.5,0){2}(10,10){$a$}{$c$}
\multiframe(82,-0.5)(10.5,0){1}(10,10){$b$}
\put(108,3){$\simeq$}
\multiframe(123,10)(10.5,0){8}(10,10){$a$}{$c$}{$d_1$}{.}{.}{.}{.}{$d_k$}
\multiframe(123,-.5)(10.5,0){1}(10,10){$b$}
\put(214,2){$\bigoplus$}
\multiframe(230,10)(10.5,0){7}(10,10){$a$}{$c$}{$d_1$}{.}{.}{.}{.}
\multiframe(230,-.5)(10.5,0){2}(10,10){$b$}{$d_k$}
\put(108,-23){$\bigoplus$}
\multiframe(123,-25)(10.5,0){7}(10,10){$a$}{$c$}{$d_1$}{.}{.}{.}{.}
\multiframe(123,-35.5)(0,-10.5){2}(10,10){$b$}{$d_k$}
\put(214,-23){$\bigoplus$}
\multiframe(230,-25)(10.5,0){6}(10,10){$a$}{$c$}{$d_1$}{.}{.}{.}
\multiframe(230,-35.5)(10.5,0){2}(10,10){$b$}{$d_{l}$}
\multiframe(230,-46)(10.5,0){1}(10,10){$d_k$}
\put(300,-50){,}
\end{picture} \normalsize

\noindent where $l=k-1$.

The combinations of
$\partial_{d_1 \ldots d_k}T_{[ab]c}$ represented by the
above diagrams are respectively denoted by
$R^T_{abcd_1...d_k}$, $H^T_{abcd_1...d_k}$,
$F^T_{abcd_1...d_k}$, and $E^T_{abcd_1...d_k}$. They are
obtained by first symmetrizing $\partial_{d_1 \ldots
d_k}T_{[ab]c}$ according to every line and then
antisymmetrizing the result according to every column. By
convention, for every symmetrization or antisymmetrization,
we divide the corresponding sum of terms by a factorial
term. For example,
\footnotesize 
\begin{picture}(35,13)(0,0)
\multiframe(0,0)(10.5,0){3}(10,10){$i$}{$j$}{$k$}
\end{picture} 
\normalsize
$\equiv
\frac{1}{3!}(Y_{ijk}+Y_{jik}+Y_{kji}+Y_{ikj}+Y_{jki}+Y_{kij})$.
The variables $R,H,F$ and $E$ form a new basis for the space
spanned by $\partial_{d_1 \ldots d_k}T_{[ab]c}$

In exactly the same way, the $(k+1)$-th order derivatives of
the ghosts $A_{ab}$ are decomposed according to,

\footnotesize
\begin{picture}(210,60)(0,-20)
\multiframe(0,0)(10.5,0){7}(10,10){$d_1$}{.}{.}{.}{.}{$d_k$}{$c$}
\put(78,2){$\bigotimes$}
\multiframe(92.5,10)(10.5,0){1}(10,10){$a$}
\multiframe(92.5,-0.5)(10.5,0){1}(10,10){$b$}
\put(108,3){$\simeq$}
\multiframe(124,10)(10.5,0){8}(10,10){$a$}{$c$}{$d_1$}{.}{.}{.}{.}{$d_k$}
\multiframe(124,-.5)(10.5,0){1}(10,10){$b$}
\put(213,2){$\bigoplus$}
\multiframe(228,10)(10.5,0){7}(10,10){$a$}{$c$}{$d_1$}{.}{.}{.}{.}
\multiframe(228,-.5)(10.5,0){1}(10,10){$b$}
\multiframe(228,-11)(10.5,0){1}(10,10){$d_k$}
\put(315,-11){.}
\end{picture}\normalsize

The diagrams of the rhs of the above decomposition
are denoted respectively by $R^A_{abcd_1...d_k}$ and
$F^A_{abcd_1...d_k}$. Note that in the case $k=0$ the above
notation is not well adapted because the diagram
\footnotesize \begin{picture}(15,35)(0,-21)
\multiframe(0,0)(0,-10.5){3}(10,10){$a$}{$b$}{$c$}
\end{picture} \normalsize
is missing. In that case, we denote the corresponding combination of
$\partial_c A_{ab}$ by $F^A_{abc}$.

For the $(k+1)$-th order derivatives of the $S_{ac}$ we have,

\footnotesize\begin{picture}(210,70)(0,-40)
\multiframe(0,0)(10.5,0){7}(10,10){$d_1$}{.}{.}{.}{.}{$d_k$}{$b$}
\put(78,2){$\bigotimes$}
\multiframe(92.5,0)(10.5,0){2}(10,10){$a$}{$c$}
\put(119.5,3){$\simeq$}
\multiframe(134,0)(10.5,0){9}(10,10){$a$}{$c$}{$b$}{$d_1$}{.}{.}{.}{.}{$d_k$}
\put(234,2){$\bigoplus$}
\multiframe(249,10)(10.5,0){8}(10,10){$a$}{$c$}{$d_1$}{.}{.}{.}{.}{$d_k$}
\multiframe(249,-.5)(10.5,0){1}(10,10){$b$}
\put(122,-23){$\bigoplus$}
\multiframe(142,-25)(10.5,0){7}(10,10){$a$}{$c$}{$d_1$}{.}{.}{.}{.}
\multiframe(142,-35.5)(10.5,0){2}(10,10){$b$}{$d_k$}
\put(225,-35.5){.}
\end{picture} \normalsize

The components of the decomposition are denoted respectively,
$L^S_{abcd_1...d_k}$, $R^S_{abcd_1...d_k}$,
$H^S_{abcd_1...d_k}$.

Finally, the $(k+2)$-th derivatives of the $C_a$ decompose
according to,

\footnotesize \begin{picture}(210,50)(0,-15)
\multiframe(0,0)(10.5,0){8}(10,10){$d_1$}{.}{.}{.}{.}{$d_k$}{$b$}{$c$}
\put(88.5,2){$\bigotimes$}
\multiframe(103,0)(10.5,0){1}(10,10){$a$}
\put(119.5,3){$\simeq$}
\multiframe(131.5,0)(10.5,0){9}(10,10){$a$}{$c$}{$b$}{$d_1$}{.}{.}{.}{.}{$d_k$}
\put(233,2){$\bigoplus$}
\multiframe(248,10)(10.5,0){8}(10,10){$a$}{$c$}{$d_1$}{.}{.}{.}{.}{$d_k$}
\multiframe(248,-.5)(10.5,0){1}(10,10){$b$}
\put(343,-.5){.}
\end{picture} \normalsize

The two different components are denoted $L^C_{abcd_1...d_k}$ and
$R^C_{abcd_1...d_k}$.

With the above definitions, an explicit calculation shows
that we have the following relations among the variables:
\begin{eqnarray}
\gamma R^T_{abcd_1...d_k}&=&3 R^A_{abcd_1...d_k} +
\frac{k+3}{2}R^S_{abcd_1...d_k}, \\
\gamma H^T_{abcd_1...d_k}&=&\frac{k+2}{2}
H^S_{abcd_1...d_k},
\\ \gamma F^T_{abcd_1...d_k}&=&3 F^A_{abcd_1...d_k},
\\ \gamma L^S_{abcd_1...d_k}&=& 6 L^C_{abcd_1...d_k},
\\ \gamma R^S_{abcd_1...d_k}&=&3 R^C_{abcd_1...d_k},
\\ \gamma E^T_{abcd_1...d_k}&=&0 \label{gammaET},
\\ \gamma F^A_{abc}&=&0.\label{gammaFA}
\end{eqnarray}

The variables now obey the relations (\ref{regroup}). We
see that up to numerical factors, the operator $\gamma$
groups in pairs the various Young diagrams of the same
symmetry type. Combinations of the variables corresponding to
Young diagrams which don't belong to a pair have a vanishing
$\gamma$-variation and are not $\gamma$-exact. From
(\ref{gammaCa}), (\ref{gammaET}) and (\ref{gammaFA}) we
conclude that the cohomology of $\gamma$ is generated by the
variables $C_a$, $F^A_{abc}$, $E^T_{abcd_1...d_k}$, the
antifields and their derivatives.

Any gauge invariant function made up of the fields and
their derivatives is thus a product of
$E^T_{abcd_1...d_k}$ which are derivatives of the curl of the
field strength. As a consequence, we note that in order to
built gauge invariant interactions we have to use at least
second order derivatives of the fields.

\section{Conclusions}
In this short paper we have calculated the cohomology of
the longitudinal exterior derivative $\gamma$ for a theory
containing generalized gauge fields represented by the
Young diagram
\footnotesize \begin{picture}(20,22)(0,0)
\multiframe(0,10)(10.5,0){2}(10,10){a}{c}
\multiframe(0,-0.5)(10.5,0){1}(10,10){b}
\end{picture}\normalsize. From this study we are able to
deduce all the possible gauge invariant interactions that can
be constructed from the fields and their derivatives. We have
shown that they are at least of second order in derivatives.

The method we use is based on the
technic of Young diagrams and the `Basic lemma' in which one
tries to associate in pairs diagrams which are of the same
symmetry type. This method can be generalized to higher
rank mixed tensors. For example, tensors corresponding to
Young diagrams with an arbitrary number of boxes in the
first column, i.e. \footnotesize \begin{picture}(20,47)(0,-24)
\multiframe(0,10)(10.5,0){2}(10,10){a}{c}
\multiframe(0,-0.5)(10.5,0){1}(10,10){b}
\put(5,-11){$\vdots$}
\multiframe(0,-24)(10.5,0){1}(10,10){d}
\end{picture}\normalsize, are treated exactly along the lines
of this article.
For tensor with more boxes on the first line, e.g.
\footnotesize\begin{picture}(55,25)(0,0)
\multiframe(0,10)(10.5,0){2}(10,10){a}{c}
\put(24.5,15){$\ldots$}
\multiframe(38.5,10)(10.5,0){1}(10,10){d}
\multiframe(0,-0.5)(10.5,0){1}(10,10){b}
\end{picture}\normalsize, one has to
impose trace conditions in order to build suitable
lagrangians.
The decomposition into irreducible components performed in
section IV then has to be made with respect to $O(n)$ instead of
$GL(n)$ to preserve the traces of tensors.
\vspace{.15cm}

{\bf Acknowledgments}: 
The authors are grateful to Marc Henneaux for suggesting the
problem and for
useful discussions. Bernard Knaepen is supported by the FNRS
(Belgium). A.G. acknowledges support from CONACyT (M\'exico)
for a postgraduate fellowship and to the Facult\'e des
Sciences, Universit\'e Libre de Bruxelles for kind hospitality
where part of this work was done.

\end{document}